\documentclass[sigconf]{acmart}
\AtBeginDocument{%
  }

\setcopyright{acmlicensed}
\copyrightyear{2024}
\acmYear{2024}
\acmDOI{XXXXXXX.XXXXXXX}

\acmConference[GenAI Evaluation KDD 2024]{August 25--26, 2024}{Barcelona, Spain}
\begin{document}

%%
%% The "title" command has an optional parameter,
%% allowing the author to define a "short title" to be used in page headers.
\title{VERA: Validation and Evaluation of Retrieval-Augmented systems}

%%
%% The "author" command and its associated commands are used to define
%% the authors and their affiliations.
%% Of note is the shared affiliation of the first two authors, and the
%% "authornote" and "authornotemark" commands
%% used to denote shared contribution to the research.
\author{Tianyu Ding}
\orcid{1234-5678-9012}
% \author{G.K.M. Tobin}
% \authornotemark[1]
% \email{webmaster@marysville-ohio.com}
\affiliation{%
  \institution{Amazon}
  \city{Arlington}
  \state{VA}
  \country{USA}
}
\email{tianyd@amazon.com}

\author{Adi Banerjee}
\affiliation{%
  \institution{Amazon}
  \city{Boston}
  \state{MA}
  \country{USA}}
\email{adibaner@amazon.com}

\author{Yunhong Li}
\affiliation{%
  \institution{Amazon}
  \city{Seattle}
  \state{WA}
  \country{USA}}
\email{yunhonl@amazon.com}

\author{Laurent Mombaerts}
\affiliation{%
  \institution{Amazon}
  \city{Luxembourg}
  \country{Luxembourg}}
\email{lmomb@amazon.lu}

\author{Tarik Borogovac}
\affiliation{%
  \institution{Amazon}
  \city{Boston}
  \state{MA}
  \country{USA}}
\email{tarikbo@amazon.com}

\author{Juan Pablo De la Cruz Weinstein}
\affiliation{%
  \institution{Amazon}
  \city{Seattle}
  \state{WA}
  \country{USA}}
\email{jcruam@amazon.com}
% \author{Anonymous Author(s)}

%%
%% By default, the full list of authors will be used in the page
%% headers. Often, this list is too long, and will overlap
%% other information printed in the page headers. This command allows
%% the author to define a more concise list
%% of authors' names for this purpose.
% \renewcommand{\shortauthors}{Ding, et al.}

%%
%% The abstract is a short summary of the work to be presented in the
%% article.
\begin{abstract}
  The increasing use of Retrieval-Augmented Generation (RAG) systems in various applications necessitates stringent protocols to ensure RAG systems' accuracy, safety, and alignment with user intentions. In this paper, we introduce VERA (Validation and Evaluation of Retrieval-Augmented Systems), a framework designed to enhance the transparency and reliability of outputs from large language models (LLMs) that utilize retrieved information. VERA improves the way we evaluate RAG systems in two important ways: (1) it introduces a cross-encoder based mechanism that encompasses a set of multidimensional metrics into a single comprehensive ranking score, addressing the challenge of prioritizing individual metrics, and (2) it employs Bootstrap statistics on LLM-based metrics across the document repository to establish confidence bounds, ensuring the repository's topical coverage and improving the overall reliability of retrieval systems. Through several use cases, we demonstrate how VERA can strengthen decision-making processes and trust in AI applications. Our findings not only contribute to the theoretical understanding of LLM-based RAG evaluation metric but also promote the practical implementation of responsible AI systems, marking a significant advancement in the development of reliable and transparent generative AI technologies.
\end{abstract}

%%
%% The code below is generated by the tool at http://dl.acm.org/ccs.cfm.
%% Please copy and paste the code instead of the example below.
%%
% \begin{CCSXML}
% <ccs2012>
%  <concept>
%   <concept_id>00000000.0000000.0000000</concept_id>
%   <concept_desc>Do Not Use This Code, Generate the Correct Terms for Your Paper</concept_desc>
%   <concept_significance>500</concept_significance>
%  </concept>
%  <concept>
%   <concept_id>00000000.00000000.00000000</concept_id>
%   <concept_desc>Do Not Use This Code, Generate the Correct Terms for Your Paper</concept_desc>
%   <concept_significance>300</concept_significance>
%  </concept>
%  <concept>
%   <concept_id>00000000.00000000.00000000</concept_id>
%   <concept_desc>Do Not Use This Code, Generate the Correct Terms for Your Paper</concept_desc>
%   <concept_significance>100</concept_significance>
%  </concept>
%  <concept>
%   <concept_id>00000000.00000000.00000000</concept_id>
%   <concept_desc>Do Not Use This Code, Generate the Correct Terms for Your Paper</concept_desc>
%   <concept_significance>100</concept_significance>
%  </concept>
% </ccs2012>
% \end{CCSXML}

% \ccsdesc{Do Not Use This Code~Generate the Correct Terms for Your Paper}
% \ccsdesc[100]{Do Not Use This Code~Generate the Correct Terms for Your Paper}

%%
%% Keywords. The author(s) should pick words that accurately describe
%% the work being presented. Separate the keywords with commas.
%% A "teaser" image appears between the author and affiliation
%% information and the body of the document, and typically spans the
%% page.
% \begin{teaserfigure}
%   \includegraphics[width=\textwidth]{sampleteaser}
%   \caption{Seattle Mariners at Spring Training, 2010.}
%   \Description{Enjoying the baseball game from the third-base
%   seats. Ichiro Suzuki preparing to bat.}
%   \label{fig:teaser}
% \end{teaserfigure}

\received{10 June 2024}
% \received[revised]{12 March 2009}
\received[accepted]{20 July 2024}

%%
%% This command processes the author and affiliation and title
%% information and builds the first part of the formatted document.
\maketitle

%% storing some style here:
% The ``\verb|acmart|'' document class can be used to prepare articles
% for any ACM publication --- conference or journal, and for any stage
% of publication, from review to final ``camera-ready'' copy, to the
% author's own version, with {\itshape very} few changes to the source.

\section{Introduction}
The integration of Retrieval-Augmented Generation (RAG) systems with Large Language Models (LLMs) has significantly advanced the field of natural language processing, particularly enhancing capabilities in areas such as open-domain question answering, fact-checking, and customer service support. These systems combine extensive data repositories with sophisticated generative capabilities to produce responses that are both relevant and informative \cite{guu2020retrieval, lewis2020retrieval}. 

Despite recent advancements, RAG systems rely on LLMs and hence face similar challenges, such as untraceable reasoning processes, supporting evidence is not provided as part of the answers, the production of "hallucinated" responses and answers that are coherent but factually incorrect or irrelevant \citep{nadeau2024benchmarking}. Furthermore, integrating these systems with additional databases presents unique challenges. Since these databases are static, they can have limited coverage on topics and can lead to outdated responses. Additionally, their large volumes can result in high computational costs.

Traditional methods for evaluating RAG systems involve extensive manual annotations and continuous human monitoring, both of which are resource-intensive \citep{zhong2022dialoglm}. To address these challenges, we have developed VERA, a scalable RAG evaluation method that utilizes LLM-based evaluation mechanisms and statistical estimators to provide annotations and evaluation tools suitable for production environments.

VERA efficiently evaluates both the retrieval and generation phases of RAG systems by measuring retrieval precision and recall to ensure optimal information retrieval and assessing the faithfulness and relevance of generated answers. Additionally, VERA enhances its evaluation by leveraging a cross-encoder that incorporates these retrieval and generation metrics to yield a single comprehensive score that can be used to rank RAG systems against each other. This singular score enables users to quickly ascertain the performance of their RAG systems, as well as make any engineering decisions related to it. For instance, whether to roll-back a deployment that caused an unforeseen change to their RAG performance \cite{upadhyay2023enhancing}.

Furthermore, VERA introduces an innovative method that utilizes bootstrap estimators to validate and assess the topicality of document repositories, which is essential for both industry and academic applications, particularly as synthetic data proliferates rapidly in the GPT era. Document repository topicality for given topics refers to the degree to which the documents stored in a repository are relevant and exclusively related to the specified topics, without contamination by unrelated or off-topic content. For example, in a repository dedicated to "Cloud Computing Sales and Marketing," topicality would be measured by the proportion of documents that focus precisely on strategies, trends, and analytics specific to selling and marketing cloud computing services, while excluding unrelated topics such as healthcare management, traditional retail marketing, or general IT infrastructure. This method evaluates the topicality of a repository by examining the relevance of the documents it contains to specific topics. For example, VERA can assess the extent to which the documents in a given repository are pertinent to a designated topic or set of queries.

\section{Related Work}
Traditionally, RAG models were evaluated based on their performance in specific downstream tasks, utilizing established metrics like EM and F1 scores for entity or sentiment classification, BERTScore and MoverScore for question answering, or accuracy for fact-checking \citep{ma2023query, wang2023instructretro, zhang2019bertscore, zhao2019moverscore, feng2024retrieval, weijia2023replug}. Tools like RALLE automate this process using task-specific metrics \citep{hoshi2023ralle}. State-of-the-art evaluation tools such as EXAM and RAGAS propose various quantifications for RAG retrieval and generation effectiveness, including context relevance and answer faithfulness \cite{saad2023ares, allardjarvis2023}. BARTScore and SelfCheckGPT focus on generation factuality and coherency. RAG evaluation also encompasses abilities indicative of its adaptability and efficiency: noise robustness, negative rejection, counterfactual robustness, and guideline adherence \cite{chen2024benchmarking, liu2023recall}.

Despite developments in evaluation metrics and tools, quantifying different aspects in RAG remains challenging due to uncertainties in inputs and outputs and limitations of existing benchmarks in capturing human preferences. The Large Model Systems Organization (LMSYS) group explores the feasibility and pros/cons of using various LLMs as automated judge for tasks in writing, math, and world knowledge \cite{zheng2024judging}. Their results reveal that strong LLM judges like GPT-4 can match both controlled and crowdsourced human preferences well, achieving over 80\% agreement, the same level of agreement between humans. The G-EVAL proposed by Microsoft Cognitive Services Research group with GPT-4 as the backbone model achieves a Spearman correlation of 0.514 with human on summarization task, along with other studies confirming GPT’s ability to achieve state-of-the-art or competitive correlation with human judgments \cite{liu2023gpteval, wang2023chatgpt}. Furthermore, several initiatives leverage LLM prompting to evaluate performance across diverse tasks such as translation, summarization, and dialogue [14]. These studies point out that LLMs offer a scalable and explainable alternative to human evaluation, which are otherwise very expensive to obtain \cite{zheng2024judging}.

Lastly, given that RAGs rely on a retrieval model to retrieve relevant documents, their performance is pegged to the efficacy of the semantic search within the retriever. As the quality of semantic search is dependent on document ingestion and chunking strategies employed, the retriever can be made more robust by a re-ranking mechanism. This is where cross-encoder models have emerged as a prominent architecture in Natural Language Processing (NLP) for tasks requiring semantic similarity assessment and textual relationship understanding \cite{crossencoderunderstand, crosencoderwhat}. These models, often leveraging transformer-based encoders like BERT, process text pairs and generate a joint embedding that encodes their semantic connection \citep{devlin2018bert}. This functionality allows for various applications, including sentence retrieval, question answering, and paraphrase detection \citep{sentencetransformers}. Cross-encoders offer advantages in efficiency compared to methods like Siamese networks, particularly for large datasets. Additionally, their ability to leverage pre-trained language models enables effective performance even with limited task-specific training data \citep{sentencetransformers}.

\section{VERA method}
VERA first systematically assesses the integrity of document repositories using LLM-based metrics, such as Retrieval Precision, Recall, Faithfulness, and Answer Relevance. It then applies advanced techniques like rank-based aggregation and bootstrapping to enhance the usability, reliability, and reproducibility of these metrics. Finally, it conducts contrastive analysis to evaluate document repository topicality \citep{wang1998cognitive}. This approach not only evaluates the relevance and accuracy of document retrieval but also ensures the integrity and thematic consistency of the information retrieved. Section 3.1 covers how VERA uses LLMs to generate integrity related metrics. Section 3.2 discusses rank-based aggregation. Section 3.3 introduces the bootstrapping technique. Section 3.4 details how contrastive analysis is used to assess document topicality.

\begin{figure}[htbp]
    \centering
    \includegraphics[width=\linewidth]{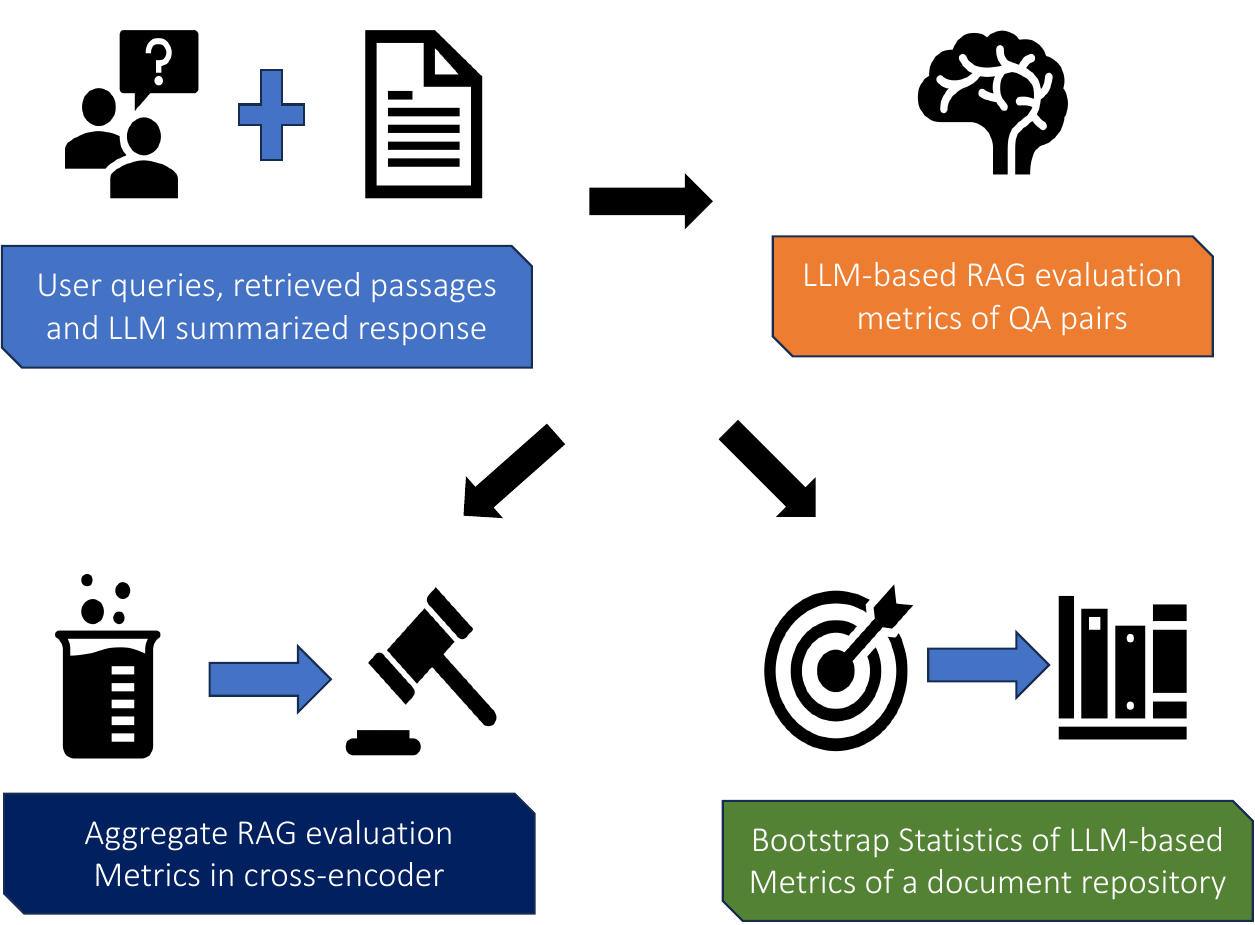}\label{fig:f1}
    \caption{Overview of VERA: VERA begins with user queries, pairing them with retrieved and LLM summarized responses from a given RAG system. These elements form the basis for the LLM-based RAG evaluation of individual question-answer pairs, ensuring that the context relevance, answer faithfulness, and answer relevance metrics are meticulously assessed. These metrics are then consolidated using a cross-encoder to generate an aggregate score, enabling users to prioritize certain metrics over others and quickly make outcome-oriented decisions for development. The process then culminates with Bootstrap Statistics, which apply LLM-based metrics across the entire document repository to establish confidence bounds and gauge the overall performance of retrieval systems. This robust evaluation pipeline is essential for maintaining high standards of precision and trustworthiness in document retrieval, particularly critical in domains where the accuracy of information is paramount.}
\end{figure}

\subsection{LLMs as Evaluators}
Recent advances in LLMs' information retrieval, understanding of nuances, and reasoning abilities have made their applications in high-stakes tasks such as system evaluations practical and feasible \cite{anthropicclaude, chung2024scaling}. VERA uses Anthropic Claude V3 Haiku through Amazon Bedrock service as the default LLM for RAG evaluations, due to Haiku's balance between cost and effectiveness. Haiku achieves competitive performance on major reasoning dataset: 75.2\% on MMLU \cite{clark2018think}, 89.2\% on ARC-Challenge \cite{zellers2019hellaswag} and 85.9\% on HellaSwag \cite{achiam2023gpt}. On each of the dataset, it has surpassed GPT-3.5 over all those three evaluation benchmark datasets. A different LLM can be chosen based on the model's merits, specific use cases, and costs.

Like existing LLM-based RAG evaluation system such as RAGAS or ARES \cite{saad2023ares}, VERA has measured the following LLM-based evaluation metrics. The prompts to create the metrics are listed in Appendix \ref{sec: prompt}.

\textbf{Faithfulness:} This metric evaluates whether answers are based solely on the provided contexts, without any fabrication. The prompt will instruct the language model to generate a binary "yes" or "no" label for each $(q, a, c)$ pair, where $q$, $a$, and $c$ represent the question, answer, and context, respectively. The faithfulness metric for a set of $(q, a, c)$ pairs is calculated as the average of all the binary labels.

\textbf{Retrieval Recall:} This metric evaluates the system’s effectiveness in fetching all relevant information related to a query from the given context, ensuring that no significant data is omitted. This metric is determined by assessing whether each piece of information in the answer is explicitly supported by the context. The retrieval recall metric is calculated based on the proportion of sentences in the answer that are classified as "[Supported by Context]", which is used in the evaluation metric prompt in Appendix \ref{sec: prompt}. This involves:

\begin{itemize}
  \item Counting the total number of sentences in the answer.
  \item Counting the number of sentences classified as "[Supported by Context]".
  \item Calculating the ratio of "[Supported by Context]" sentences to the total number of sentences.
\end{itemize}

\textbf{Retrieval Precision:} This metrics assesses the system’s ability to focus on and retrieve the most relevant parts of the context in response to a query, minimizing the inclusion of irrelevant content. High precision ensures that the model considers only the information that is directly pertinent to the question. The retrieval precision metric is calculated by evaluating the relevance of sentences extracted from the context using LLM. This involves: 
\begin{itemize}
  \item Extracting sentences from the context that directly support the answer to the question.
  \item Ensuring that sentences are not altered when extracted.
  \item Returning "Insufficient Information" if no relevant sentences are found or if the context does not provide enough information to answer the question.
  \item Measuring the similarity between the extracted sentences and the context using embedding models. In this work, we used Amazon Titan embedding model \cite{amazonbedrockdocumentation}.
  \item Calculating the precision based on the ratio of relevant sentences to the total number of sentences in the context.
\end{itemize}

\textbf{Answer Relevance:} This metrics evaluates whether the generated response directly addresses the given question, ensuring alignment with both the query and the retrieved context. This metric penalizes responses that are incomplete, redundant, or contain unnecessary information, providing a score that ranges from 0 to 1, with 1 being the highest level of relevance. The answer relevance metric is calculated through the following steps:

\begin{itemize}
  \item For each generated answer, multiple questions are generated to assess the alignment of the answer with the query.
  \item The similarity between the generated questions and the original question is measured using embeddings. This involves:
    \begin{enumerate}
      \item Embedding the original question and the generated questions using an embedding model. In this work, we used Amazon Titan embedding model \cite{amazonbedrockdocumentation}.
      \item Calculating the cosine similarity between the original question embedding and each generated question embedding.
    \end{enumerate}
  \item The final score is computed as the mean cosine similarity across all generated questions for each answer, reflecting the degree of relevance.
\end{itemize}

\subsection{Consolidation of Multi-Dimensional Evaluation Metrics}
The concept of consolidating evaluation metrics into a single comprehensive score involves integrating the utilities of each metric, allowing users to make informed decisions despite the inherent fluctuations in these metrics. Appropriate consolidation eases the burden of users having to parse through multiple metrics to then make a decision based on the outcome - which would improve iteration speed in a development cycle. Furthermore, given that each of these multi-dimensional metrics has its nuances, the question of how to prioritize certain metrics over others arises (e.g. does a system with higher faithfulness and lower relevance outperform the system with lower faithfulness and higher relevance). This would assist users to swiftly take action during regression testing, to make decisions on whether to roll-back a deployment or not.

Traditional techniques like simple aggregation or rank fusion often suffer from compensatory effects and lack clarity, as they obscure the subtleties of individual metrics \cite{arun2017integrating, safjanlambdamart}.

To address these challenges, VERA utilizes cross-encoder models that leverage a cross-attention mechanism for a more precise evaluation of document relevance. Traditional cross-encoder models are effective at highlighting the most relevant text segments within large texts, based on capturing semantic relationships between words and phrases. It generates a relevance score for every question-answer pair, enabling an effective comparison and ranking of these pairs. Formally, for a user-input question q and an answer a, the logit score $\sigma$ is determined as:
  \[\sigma(q,a)=CE([CLS]\;q\;[SEP]\;a\;[SEP])\; \times W\]
where CE is the cross-encoder, CLS and SEP are special tokens to represent the classifier token and the separator token, and W is a learned matrix that represents the relationship between the query and answer representations \cite{upadhyay2023enhancing}. 

Recently, effective multi-dimensional retrieval models are typically implemented by performing a first-stage retrieval (to efficiently identify a subset of relevant documents from a corpus); and a second-stage re-ranking on this subset (where additional dimensions of relevance may be considered) \cite{upadhyay2023enhancing}. An example of this to conduct the first stage retrieval using the BM-25 algorithm (which is a ranking algorithm that determines a document's relevance to a given query and ranks documents based on their relevance scores). After this, the second-stage re-ranker modifies the architecture of existing cross-encoders, whereby the BM-25 score obtained in the first-stage retrieval is fed as an input token to the cross-encoder. Mathematically, this is represented by:
  \[\sigma(q,a)=CE([CLS]\;q\;[SEP]\;BM25\;[SEP]\;a\;[SEP])\; \times W\] 

In this paper, we follow a similar process to incorporate additional dimensions of relevance into a cross-encoder to re-rank evaluation records against each other. However, instead of manipulating the input structure of the cross-encoder, we integrate additional "relevance statements" into each question-answer pair that is fed into the cross-encoder \cite{upadhyay2023enhancing}. These relevance statements pertain to texts related to the utility of each of the multi-dimensional evaluation metrics, as well as their actual scores. As shown in \cite{upadhyay2023enhancing}, this exercise yields 4-5 percent improvement in Mean Average Precision, Normalized Discounted Cumulative Gain and Mean Reciprocal Rank metrics against baseline cross-encoder models.

The process involves two key steps: first, enhancing the input text with additional relevant information mentioned above, and second, providing the queries and “enhanced” answer as inputs to the pretrained cross-encoder to obtain the final aggregated score (which can be used to rank these records against each other) \cite{upadhyay2023enhancing}. This structured approach ensures a thorough and nuanced assessment of document relevance.
 
\textbf{Text Enhancement:}
The cross-encoder requires an input of an input and output text. Within VERA, the input text will be the user query input to a RAG system; and the output text will be an enriched answer encapsulating the original response from the RAG system, together with supplementary information about each evaluated metric’s utility as well as their scores. For instance, a question-answer pair (q,a) obtains a faithfulness score of 0.7 and its enriched answer $\bar{a}$ is generated by appending the original response from the RAG system with the following text:

"\textit{Faithfulness measures the factual consistency of the generated answer against the given context. It is considered faithful if all the claims that are made in the answer can be inferred from the given context. It is measured between 0 and 1; where a lower score is given to answers consisting of claims that are not in the context; and a higher score indicates that the answer is using information from the contexts. For the given question, context and answer, the faithfulness score is 0.7.}"

\textbf{Cross-Encoder Ranking:}
Once the text enhancement step is done, VERA feeds in the question and the enhanced answer into the ms-marco-MiniLM-L-12-v2 (top cross-encoder model on MTEB Leaderboard). Formally, for a user-input question q and enriched answer $\bar{a}$, the logit score $\sigma$ is determined as:
  \[\sigma(q,\bar{a})=CE([CLS]\;q\;[SEP]\;\bar{a}\;[SEP])\;.\;W\]
As this cross-encoder model was trained to learn logit values, it can be normalized to a value between 0 and 1 by taking the expit. However, this paper will present the results as logit scores.

\subsection{Bootstrap LLM-based RAG Evaluation Metrics}
\label{subsec: bootstrap}
Evaluating RAG systems requires measuring retrieval precision, recall, faithfulness, and relevance. However, these metrics can vary due to LLMs' stochasticity, reasoning limitations, and document repository topicality. To address this, we used bootstrapping on pre-computed metric values. This approach enhances result reliability and reproducibility by providing a robust statistical framework to analyze metric variability and distribution, while also supporting document repository topicality assessment for specific content types \cite{hesterberg2011bootstrap}.

LLMs can produce varying outputs due to factors like random seed values, causing traditional evaluation to capture only a snapshot of this variability and potentially misleading performance conclusions. By applying bootstrap directly on the metric values, we can simulate multiple runs of model evaluations, capturing a broader spectrum of possible outcomes and thus providing a more comprehensive picture of system performance.

Bootstrapping metric values allows for repeated sampling from a set of observed metric computations, essentially creating numerous virtual evaluation scenarios. The bootstrapping metric values computation are identical for all metrics. 

Given a known metric $M$: Firstly, compute its values for a document repository dataset $D=\{d_1,d_2,...,d_n\}$ as $M(d_i)$ for each document $d_i$. This results in a set of metric values $ M = \{m_1,m_2,...,m_n\}$. Then, for each metric $M$, generate $B$ bootstrap samples. Each sample $s$ is created by randomly selecting metric values from $M$ with replacement. Each bootstrap sample for metric $M$ can be represented as $ M_s  = \{m_1^s,m_2^s,...,m_n^s\} $. For each bootstrap sample, compute the desired statistics, such as the sample mean and variance as below:

\begin{itemize}
  \item Estimates the Mean and Variability: Provides a statistically robust way to estimate the mean and variance of performance metrics, incorporating the inherent randomness of LLM outputs. The mean $\bar{M}$ of the bootstrap samples is estimated as:
  \[\bar{M}=\frac{1}{B}\sum_{s=1}^{B}\bar{m}_s\]
  And the variance $\sigma^2(M)$ is:
  \[\sigma^2(M)=\frac{1}{B-1}\sum_{s=1}^{B}(\bar{m}_s - \bar{M})^2\]
  \item Confidence intervals: Can be derived from the percentiles of the bootstrap distribution, typically the $2.5th$ and $97.5th$ percentiles for a 95\% confidence interval.

\end{itemize}

\textbf{Bootstrapping Size B versus Sample Size n:}
There is no strict universal rule for the optimal bootstrapping size relative to the sample size. However, bootstrapping tends to work well when the sample size (n) is at least 30 and ideally 50 or more for accurate estimations of standard errors and confidence intervals \cite{tibshirani1993introduction, wu1986jackknife}. Bootstrapping size is recommended to be at least 1000 and 5000 or more for stable convergence and complex statistics. With larger sample size, a smaller bootstrapping size is possible with similar accuracy. It is recommended to monitor the convergence of standard errors or other statistics as $B$ increases, to determine the optimal bootstrapping size for each use case.

\textbf{Unbiased Estimator:}
The bootstrap estimator serves as an unbiased estimator for metrics based on LLMs, effectively estimating the expectation of the original estimator and its bootstrap distribution. Detailed assumptions and mathematical derivations supporting this conclusion are outlined in Appendix \ref{sec:proof}.

\subsection{Evaluating Document Repository Topicality Using Contrastive Query Analysis}
Document repositories often contain diverse content, leading to high entropy for domain-specific information retrieval and making it challenging to ascertain the repository's thematic topicality, especially in specialized industry domains. To address this, we implement a contrastive analysis framework, differentiating responses to topic perfect relevant queries (positive instances) from responses to unrelated queries (negative controls). Within the framework, we have proposed a bootstrap estimation approach that provides a structured statistical analysis to evaluate the repository’s thematic consistency.

The approach involves several key steps with the idea ignited from contrastive learning \cite{chuang2020debiased}:

\begin{itemize}
\item Query Generation: Develop two distinct sets of queries. Positive queries set are relevant to a specific domain of interest, and negative queries are deliberately chosen to be unrelated to that domain. 
\item Retrieval and Evaluation: Utilize a large language model (LLM) or a similar retrieval system to fetch and evaluate responses for each query. Evaluation metrics such as Retrieval Precision, Recall, Faithfulness, and Answer Relevance are calculated to assess the quality and relevance of the responses. 
\item Bootstrap Statistics: Apply bootstrap sampling techniques to each evaluation metrics. This involves generating numerous subsamples from the collected metrics and computing statistical measures (e.g., mean, variance) for these samples to analyze the data robustly. 
\item Comparative Analysis: Compare the distributions of these bootstrap statistics between the positive and negative query sets. This step quantitatively assesses the repository’s content alignment with the domain of interest and identifies any significant disparities in content handling between relevant and irrelevant queries.
\end{itemize}

\section{Experiments}
In this section, we present the models and data used by VERA. VERA uses both public and proprietary datasets to ensure a comprehensive analysis. We utilize the open-source MS MARCO in TREC 2023 Deep Learning Track for a general knowledge \cite{lawrie2024overview}. Simultaneously, we incorporate proprietary datasets tailored to AWS sales and marketing domain-specific evaluations, reflecting the unique challenges and requirements of different industry sectors. This combination allows us to assess the general applicability and targeted performance of our RAG systems, facilitating a thorough understanding of their capabilities and areas for optimization in real-world scenarios.

\subsection{Models}
For domain-specific synthetic data generation, we employ Anthropic V3 Haiku to create high-quality synthetic queries and responses tailored for our experimental needs. This model's advanced generative capabilities ensure that the synthetic datasets are both diverse and closely aligned with the task-specific requirements. For the evaluation of responses, we utilize Anthropic V3 Sonnet, which serves as our LLM judge. The examples of synthetic generation prompts, evaluation prompts and RAG summarization prompt using Llama 3 supported in POE Web UI \cite{thorne2018fever} are in Appendix \ref{sec: prompt}.

In our experiments, we compared the performance of multiple RAG systems by pairing different combinations of LLMs—specifically Anthropic Haiku and Llama3—with a selection of advanced retrievers. The retrievers we've chosen include e5-mistral-7b-instruct, titan-embedding-text-G1, and bge-large-en-v1.5, all of which are recognized as top models on the MTEB leaderboard, indicative of their superior performance and capability in handling complex retrieval tasks \cite{wang2023improving,wang2022text,amazonbedrockdocumentation,xiao2023c}. This diverse combination of cutting-edge LLMs and retrievers allows us to thoroughly assess and contrast the strengths and limitations of different RAG configurations in producing relevant and accurate responses.

\subsection{Datasets}
\label{subsec:dataset}
The TREC 2023 Deep Learning Track emphasizes enhancing information retrieval with large-scale datasets suitable for deep learning, focusing on passage and document ranking tasks. It utilizes the MS MARCO dataset to analyze and develop effective retrieval and reranking systems in real-world scenarios. In this research, we'll focus on using the smaller passage ranking data from the TREC 2023 Deep Learning Track for experimental purpose. For the purpose of our experiments, we have used all 887 unique perfectly relevant query-passage pairs (score=3) from "2021.qrels.docs.final.txt" and 500 randomly sampled irrelevant query-passage pairs (score=0).

Additionally, we have generated 400 passages related to cloud computation sales and marketing topic and 100 passages related to basketball topic. Then, we have created 200 queries about cloud computation sales and marketing, 200 queries about basketball and 200 random queries not related to both topics.

\section{Results \& Analysis}
\subsection{VERA LLM-Based RAG Evaluation Metrics}
In this experiment, we evaluate the performance of several RAG systems by comparing perfectly relevant and irrelevant query-passages pairs on faithfulness, answer relevance, retrieval recall, retrieval precision, as well as the logit values returned by the cross-encoder when performing the aggregation step ("Agg" column in tables) . Findings are presented in Table~\ref{table:1} and Appendix Table~\ref{table:3}. All the metric values in both tables are reported as mean.

We feed the top 5 retrieved passages for a given query into a LLM to generate the final summarized response. In this experiment, the dataset labels "PR" and "IR" stand for "Perfectly Relevant" and "Irrelevant," respectively. To make the results more deterministic and less affected by the randomness inherent in LLMs, we have implemented the following settings: $temperature=0$; $top \: p=0.01$. The metric values in both tables are mean of all queries' results. 

\begin{table*}[h!]
\begin{center}
% \small
\caption{VERA LLM-Based RAG Evaluation Metrics on 500 Perfectly Relevant MS MARCO TREC 2023 Query-Passage Pairs}
\begin{tabular}{p{5cm} p{1.5cm} p{1.5cm} p{1.5cm} p{1.5cm} p{1.5cm} p{1.5cm}} 
 \toprule
 Models & Dataset & Faithfulness & Relevance & Recall & Precision & Aggregate Logit Score \\ 
 \midrule
 Llama3 + e5-mistral-7b-instruct & PR & 0.94 & 0.87 & 0.76 & 0.68 & 8.72 \\ 
 Llama3 + titan-embedding-text-G1 & PR & 0.93 & 0.81 & 0.74 & 0.63 & 8.69 \\
 Llama3 + bge-large-en-v1.5 & PR & 0.94 & 0.83 & 0.77 & 0.64 & 8.70 \\
 Haiku + e5-mistral-7b-instruct & PR &  0.95 & 0.85 & 0.75 & 0.65 & 8.48 \\
 Haiku + titan-embedding-text-G1 & PR & 0.93 & 0.81 & 0.73 & 0.64 & 8.59 \\ 
 Haiku + bge-large-en-v1.5 & PR & 0.94 & 0.82 & 0.74 & 0.63 & 8.46  \\ 
T-5 Flan + all-MiniLM-L6-v2 & PR & 0.82 & 0.53 & 0.61 & 0.50 & 6.11 \\
\bottomrule
\end{tabular}
\label{table:1}
\end{center}
\end{table*}

\begin{table*}[h!]
% \small
\begin{center}
\caption{VERA LLM-Based RAG Evaluation Metrics on 500 Irrelevant MS MARCO TREC 2023 Query-Passage Pairs}
\begin{tabular}{p{5cm} p{1.5cm} p{1.5cm} p{1.5cm} p{1.5cm} p{1.5cm} p{1.5cm}} 
 \toprule
 Models & Dataset & Faithfulness & Relevance & Recall & Precision & Agg \\
 \midrule
 Llama3 + e5-mistral-7b-instruct & IR & 0.94 & 0.20 & 0.10 & 0.12 & 6.45\\ 
 Llama3 + titan-embedding-text-G1 & IR & 0.94 & 0.13 & 0.09 & 0.11 & 6.23\\
 Llama3 + bge-large-en-v1.5 & IR & 0.95 & 0.12 & 0.11 & 0.14 & 6.39\\
 Haiku + e5-mistral-7b-instruct & IR & 1.0 & 0.08 & 0.03 & 0.12 & 6.38\\
 Haiku + titan-embedding-text-G1 & IR & 1.0 & 0.29 & 0.22 & 0.12 & 6.26\\ 
 Haiku + bge-large-en-v1.5 & IR & 1.0 & 0.21 & 0.10 & 0.13 & 6.55\\ 
 T-5 Flan + all-MiniLM-L6-v2 & IR & 0.87 & 0.12 & 0.02 & 0.05 & 3.81 \\
\bottomrule
\end{tabular}
\label{table:2}
\end{center}
\end{table*}

The experimental results demonstrate that Llama3, a powerful open-source LLM, performs comparably to established models like Anthropic's Claude V3 Haiku. In Table~\ref{table:1}, these models manage effective fact-checking and capture semantic relationships well, as indicated by high faithfulness and relevance scores. Additionally, the retrieval recall and precision are reasonably high, suggesting that the models retrieve most relevant information accurately. In the opposite way, the low precision score in Appendix Table~\ref{table:3} may suggest that the queries either fall outside the scope of the covered topics in the knowledge base, or that the topics within the knowledge base are too varied relative to the generality of the queries. The Agg-Logit scores in the comparison between different model configurations highlight the nuanced performance differences across various metrics.

Lastly, the performance of these powerful LLMs and embedding models have been compared to that of a weaker LLM (T-5 FLAN Base) and embedding model (all-MiniLM-L6-v2) as a "baseline", to highlight the differences in the evaluation metrics. As seen in both Table~\ref{table:1} and Table~\ref{table:2}, the individual evaluation metrics as well as the Agg-Logit scores are consistently lower when using a weaker LLM + embedding model, regardless of the scenario of evaluating against a perfectly relevant or irrelevant dataset.

\subsection{Bootstrap Metrics for Document Repository Topicality Analysis}
As outlined in section \ref{subsec: bootstrap}, we utilized bootstrap statistics to analyze a synthetic dataset described in section \ref{subsec:dataset} and the results are in Table \ref{table:3}. We used bootstrap sampling with replacement on the synthetic query sets and the overall passage set. In the synthetic query sets, we have 200 synthetic queries in each set and we labeled them in Tale \ref{table:3} as "Sales", "Basketball" and "Random" based on the topics. This approach enabled us to calculate critical statistical measures like the mean and variance, providing a robust foundation for assessing the thematic topicality of the data repository. We used sample size 50 and bootstrapping size 500 to ensure fairly stable convergence of the statistics for each metric and each query set. This comparative analysis helps in quantifying the document repository's content topicality to distinguish and accurately process content relevant to its designated domain. 

\begin{table*}[h!]
\begin{center}
% \small
\caption{VERA Bootstrap Statistics on Three Comparative QuerySet}
\begin{tabular}{p{5cm} p{2cm} p{2cm} p{2cm} p{2cm} p{2cm}} 
 \toprule
 Models & QuerySet & Faithfulness & Relevance & Recall & Precision \\
% \hline
\midrule
 Llama3 + e5-mistral-7b-instruct & Sales & 0.93±0.03 & 0.71±0.04 & 0.61±0.07 & 0.54±0.09 \\ 
 Llama3 + titan-embedding-text-G1 & Sales & 0.94±0.03 & 0.70±0.04 & 0.62±0.08 & 0.55±0.08 \\
 Llama3 + bge-large-en-v1.5 & Sales & 0.93±0.03 & 0.70±0.05 & 0.60±0.07 & 0.53±0.10 \\
 Haiku + e5-mistral-7b-instruct & Sales & 0.93±0.02 & 0.72±0.05 & 0.62±0.07 & 0.55±0.06 \\
 Haiku + titan-embedding-text-G1 & Sales & 0.94±0.03 & 0.71±0.04 & 0.63±0.07 & 0.56±0.07 \\
 Haiku + bge-large-en-v1.5 & Sales & 0.93±0.02 & 0.70±0.05 & 0.61±0.08 & 0.56±0.09 \\ 
 \midrule
  Llama3 + e5-mistral-7b-instruct &  Basketball & 0.94±0.03 & 0.67±0.05 & 0.56±0.07 & 0.43±0.09 \\ 
 Llama3 + titan-embedding-text-G1 & Basketball & 0.93±0.03 & 0.66±0.06 & 0.53±0.07 & 0.42±0.08 \\
 Llama3 + bge-large-en-v1.5 & Basketball & 0.94±0.03 & 0.66±0.05 & 0.54±0.08 & 0.45±0.10 \\
 Haiku + e5-mistral-7b-instruct & Basketball & 0.95±0.02 & 0.66±0.06 & 0.53±0.09 & 0.43±0.09 \\
 Haiku + titan-embedding-text-G1 & Basketball & 0.94±0.03 & 0.65±0.06 & 0.52±0.08 & 0.45±0.08 \\
 Haiku + bge-large-en-v1.5 & Basketball & 0.93±0.02 & 0.66±0.05 & 0.54±0.08 & 0.44±0.08 \\ 
 \midrule
 Llama3 + e5-mistral-7b-instruct & Random & 0.93±0.02 & 0.23±0.04 & 0.13±0.05 & 0.09±0.05 \\ 
 Llama3 + titan-embedding-text-G1 & Random & 0.93±0.03 & 0.21±0.05 & 0.15±0.04 & 0.10±0.04 \\
 Llama3 + bge-large-en-v1.5 & Random & 0.94±0.03 & 0.16±0.05 & 0.11±0.04 & 0.08±0.05 \\
 Haiku + e5-mistral-7b-instruct & Random & 0.94±0.02 & 0.24±0.06 & 0.12±0.04 & 0.10±0.04 \\
 Haiku + titan-embedding-text-G1 & Random & 0.92±0.03 & 0.22±0.06 & 0.14±0.05 & 0.09±0.05 \\
 Haiku + bge-large-en-v1.5 & Random & 0.93±0.03 & 0.17±0.05 & 0.14±0.04 & 0.08±0.05 \\ 
\bottomrule
\end{tabular}
\label{table:3}
\end{center}
\end{table*}

In our study, the use of bootstrap statistics enabled us to compute the mean and confidence intervals for each performance metric across three different synthetic query sets on the same document repository. This comparison revealed notable differences in retrieval-related metrics among the query sets regarding different topics. The "Sales" query set results are with higher values in recall, precision, and relevance as the majority (80\%) of the synthetic passage set is related cloud computation sales and marketing data. As comparison, the "Basketball" query set results are much higher than the "Random" query set and fairly lower than the "Sales" query set, which is within expectation and validated the effectiveness of bootstrapping approach to evaluate the document repository topicality.

\section{Conclusion}
In this paper, we introduced VERA, a framework tailored for evaluating Retrieval-Augmented Generation (RAG) systems. By generating LLM-based RAG evaluation metrics such as faithfulness, answer relevance, retrieval precision and retrieval recall, VERA can help evaluate and validate if the response from RAG based AI assistant is accurate or not. This framework boosts the reliability and transparency of RAG systems and build the trust in AI applications for users.

Our findings demonstrate VERA's capacity to enhance decision-making processes effectively. The framework has been applied across several use cases, illustrating its ability to adapt to dynamic environments and maintain the integrity of data repositories. This adaptability makes VERA an important tool in the landscape of modern AI technologies, where the accuracy and relevance of information are paramount.

Looking forward, we aim to further refine the metrics within VERA and expand its applicability to a broader range of domains and languages. Continuous advancements in VERA's methodologies will allow it to keep pace with rapid technological developments in AI. This evolution will ensure that emerging AI technologies are leveraged responsibly, maximizing their potential benefits for society.

\section{Limitations}
This paper presents several limitations that could potentially impact the comprehensiveness and applicability of its conclusions. Firstly, the analysis omits scenarios involving fine-tuned LLMs. Potential enhancements or specific use-case efficiencies brought by fine-tuned models might not be fully captured. This omission could lead to an incomplete understanding of the capabilities and limitations of the models under different experimental conditions. And the exclusion of some top proprietary LLMs in our experiments, such as OpenAI models, limits the evaluation's scope and understanding of our selected models' performance against the best available options.

Secondly, our study does not address multilingual capabilities. The focus solely on English-language tasks may limit the generalizability of our conclusions to multilingual applications. This oversight could restrict the utility of our findings for developers and researchers working on systems intended for diverse linguistic environments, potentially overlooking significant performance variations across languages. 

Thirdly, although our bootstrap estimators offer a more convincing assessment of the content complexity within a document repository, they are computationally intensive. We aim to develop a theoretically grounded, cost-effective measurement approach by constructing a pseudo-bootstrap strategy. This strategy will utilize pre-calculated evaluation metrics instead of relying on bootstrap sampling from queries.

Lastly, our study did not analyze all popular publicly available benchmarks such as the Knowledge Intensive Language Tasks (KILT) benchmark, which could have provided additional insights into the models' capabilities in retrieving, reasoning, and synthesizing information from knowledge bases in real-world scenarios \cite{petroni2020kilt,kwiatkowski2019natural,yang2018hotpotqa,thorne2018fever}.

%%
%% The next two lines define the bibliography style to be used, and
%% the bibliography file.
\bibliographystyle{ACM-Reference-Format}
\bibliography{references}

%%
%% If your work has an appendix, this is the place to put it.
\section{Appendix}

\subsection{Prompt Examples}
\label{sec: prompt}

\noindent 
\rule{\linewidth}{0.3mm}
\begin{center}
\textbf{Prompt of Faithfulness Metric}
\label{prompt:faithfulness}
\end{center}
\noindent 
\rule{\linewidth}{0.3mm}

\noindent
Consider the given context and following statements, then determine whether they are supported by the information present in the context. Provide a brief explanation for each statement before arriving at the verdict (Yes/No). Provide a final verdict for each statement in order at the end in the given format. Do not deviate from the specified format.

\bigskip
\noindent
\textbf{Context:} \\
Emma is a graduate student specializing in marine biology at Coastal University. She has a keen interest in coral reefs and is conducting her thesis on coral bleaching. Emma attends several seminars related to marine ecosystems and is actively involved in field research in the nearby coral reefs. She often collaborates with other researchers to publish her findings.

\bigskip
\noindent
\textbf{Statements:}

\begin{enumerate}
    \item Emma is studying mechanical engineering.
    
    \item Emma is working on a project related to coral reefs.
    
    \item Emma often attends computer science workshops.
    
    \item Emma collaborates with other researchers.
    
    \item Emma's research focuses on marine ecosystems.
\end{enumerate}

\bigskip
\noindent
\textbf{Answer:}

\bigskip
\noindent
1. Emma is studying mechanical engineering.

\noindent
\textbf{Explanation:} The context specifies that Emma is specializing in marine biology, not mechanical engineering. There is no information suggesting she is studying mechanical engineering.

\noindent
\textbf{Verdict:} No.

\bigskip
\noindent
2. Emma is working on a project related to coral reefs.

\noindent
\textbf{Explanation:} It is mentioned that Emma is conducting her thesis on coral bleaching, which directly relates to coral reefs.

\noindent
\textbf{Verdict:} Yes.

\bigskip
\noindent
3. Emma often attends computer science workshops.

\noindent
\textbf{Explanation:} The context only mentions that Emma attends seminars related to marine ecosystems. There is no mention of her attending computer science workshops.

\noindent
\textbf{Verdict:} No.

\bigskip
\noindent
4. Emma collaborates with other researchers.

\noindent
\textbf{Explanation:} The context explicitly states that Emma often collaborates with other researchers to publish her findings.

\noindent
\textbf{Verdict:} Yes.

\bigskip
\noindent
5. Emma's research focuses on marine ecosystems.

\noindent
\textbf{Explanation:} Emma's interest in coral reefs and her participation in seminars related to marine ecosystems suggest that her research focuses on marine ecosystems.

\noindent
\textbf{Verdict:} Yes.

\bigskip
\noindent
\textbf{Final verdict for each statement in order:} No. Yes. No. Yes. Yes.

\bigskip
\noindent
\textbf{Context:} \{context\}

\noindent
\textbf{Statements:} \{statements\}

\noindent
\textbf{Answer:}

\noindent 
\rule{\linewidth}{0.3mm}

\noindent 
\rule{\linewidth}{0.3mm}
\begin{center}
\textbf{Prompt of Retrieval Recall Metrics}
\end{center}
\noindent 
\rule{\linewidth}{0.3mm}

\noindent
Task: Given a context and an answer, analyze each sentence in the answer and classify whether the sentence is supported by the given context or not. Think in steps and reason before coming to a conclusion.

\bigskip
\noindent
\textbf{Guidelines:}

\begin{enumerate}
    \item Read each sentence in the answer carefully.
    \item Compare the sentence with the context to see if the information is explicitly mentioned.
    \item Classify each sentence as either [Supported by Context] or [Not Supported by Context].
    \item Provide a brief reasoning for your classification.
\end{enumerate}

\bigskip
\noindent
\textbf{Example 1:}

\bigskip
\noindent
\textbf{Context:}

\noindent
Isaac Newton (25 December 1642 – 20 March 1726/27) was an English mathematician, physicist, astronomer, alchemist, and author. He is widely recognized as one of the most influential scientists of all time and a key figure in the scientific revolution. His book "Philosophiæ Naturalis Principia Mathematica," first published in 1687, laid the foundations of classical mechanics. Newton made seminal contributions to optics and shares credit with Gottfried Wilhelm Leibniz for developing calculus.

\bigskip
\noindent
\textbf{Answer:}

\noindent
Isaac Newton was an English mathematician, physicist, and astronomer. He is known for writing "Philosophiæ Naturalis Principia Mathematica." Newton invented calculus independently of Leibniz.

\bigskip
\noindent
\textbf{Classification:}

\begin{enumerate}
    \item Isaac Newton was an English mathematician, physicist, and astronomer. This information is in the context. So [Supported by Context]
    \item He is known for writing "Philosophiæ Naturalis Principia Mathematica." This is explicitly mentioned in the context. So [Supported by Context]
    \item Newton invented calculus independently of Leibniz. The context mentions Newton shares credit with Leibniz for developing calculus but does not state he did it independently. So [Not Supported by Context]
\end{enumerate}

\bigskip
\noindent
\textbf{Example 2:}

\bigskip
\noindent
\textbf{Context:}

\noindent
Marie Curie (7 November 1867 – 4 July 1934) was a Polish and naturalized-French physicist and chemist who conducted pioneering research on radioactivity. She was the first woman to win a Nobel Prize, the only woman to win the Nobel prize twice, and the only person to win the Nobel Prize in two different scientific fields. Her achievements include the development of the theory of radioactivity, techniques for isolating radioactive isotopes, and the discovery of two elements, polonium and radium.

\bigskip
\noindent
\textbf{Answer:}

\noindent
Marie Curie was a Polish physicist who won the Nobel Prize twice. She discovered the elements polonium and radium. Curie was the first person to win Nobel Prizes in two different fields.

\bigskip
\noindent
\textbf{Classification:}

\begin{enumerate}
    \item Marie Curie was a Polish physicist who won the Nobel Prize twice. This is explicitly mentioned in the context. So [Supported by Context]
    \item She discovered the elements polonium and radium. This is explicitly mentioned in the context. So [Supported by Context]
    \item Curie was the first person to win Nobel Prizes in two different fields. This is explicitly mentioned in the context. So [Supported by Context]
\end{enumerate}

\bigskip
\noindent
\textbf{Context:} \{context\}

\bigskip
\noindent
\textbf{Answer:} \{ground\_truth\}

\bigskip
\noindent
\textbf{Classification:}

\noindent 
\rule{\linewidth}{0.3mm}

\noindent 
\rule{\linewidth}{0.3mm}
\begin{center}
\textbf{Prompt of Retrieval Precision Metric}
\end{center}
\noindent 
\rule{\linewidth}{0.3mm}

\noindent
Task: Evaluate whether the provided context can answer the given question by extracting relevant sentences. Follow these guidelines:

\begin{enumerate}
    \item \textbf{Extract Sentences}: Identify and extract sentences from the context that directly support an answer to the question.
    
    \item \textbf{No Modifications}: Do not alter the sentences when extracting them.
    
    \item \textbf{Insufficient Information}: If no relevant sentences are found or if the context does not provide enough information to answer the question, return "Insufficient Information".
\end{enumerate}

\bigskip
\noindent
\textbf{Examples:}

\bigskip
\noindent
\textbf{Example 1:}

\noindent
\textbf{Question:} What causes the tides to rise and fall?

\noindent
\textbf{Context:} The gravitational pull of the moon and the sun causes the tides to rise and fall. The moon's gravity has a greater effect because it is closer to the Earth, creating high and low tides. The sun also plays a role, but to a lesser extent.

\noindent
\textbf{Candidate Sentences:}

\begin{itemize}
    \item The gravitational pull of the moon and the sun causes the tides to rise and fall.
    \item The moon's gravity has a greater effect because it is closer to the Earth, creating high and low tides.
\end{itemize}

\bigskip
\noindent
\textbf{Example 2:}

\noindent
\textbf{Question:} Who discovered penicillin?

\noindent
\textbf{Context:} Penicillin was discovered by Alexander Fleming in 1928. He noticed that a mold called \textit{Penicillium notatum} had killed a staphylococcus bacterium in a petri dish. This discovery led to the development of antibiotics, which have saved countless lives.

\noindent
\textbf{Candidate Sentences:}

\begin{itemize}
    \item Penicillin was discovered by Alexander Fleming in 1928.
    \item He noticed that a mold called \textit{Penicillium notatum} had killed a staphylococcus bacterium in a petri dish.
\end{itemize}

\bigskip
\noindent
\textbf{Example 3:}

\noindent
\textbf{Question:} What is the capital of Atlantis?

\noindent
\textbf{Context:} Many myths surround the lost city of Atlantis, but no concrete evidence has ever been found to confirm its existence. Some legends suggest it was a powerful civilization located in the Atlantic Ocean, but its exact location and details remain unknown.

\noindent
\textbf{Candidate Sentences:} \\
Insufficient Information

\bigskip
\noindent
\textbf{Question:} \{question\}

\noindent
\textbf{Context:}

\noindent
\{context\}

\bigskip
\noindent
\textbf{Candidate Sentences:}

\noindent 
\rule{\linewidth}{0.3mm}

\noindent 
\rule{\linewidth}{0.3mm}
\begin{center}
\textbf{Question Generation Prompt for Answer Relevance Metric}
\end{center}
\noindent 
\rule{\linewidth}{0.3mm}

\noindent
Task: Generate a question based on the given answer. The question should be specific, clear, and directly related to the information provided in the answer.

\noindent
Guidelines:

\begin{enumerate}
    \item \textbf{Identify Key Information}: Carefully read the given answer to identify the key pieces of information. These may include dates, times, locations, events, people, etc.

    \item \textbf{Formulate the Question}: Create a question that specifically asks about the key information you identified in the answer. The question should be comprehensive and direct, ensuring it covers all the important details provided in the answer.

    \item \textbf{Ensure Clarity and Specificity}: The question should be clear and specific, leaving no ambiguity about what information it seeks. It should be framed in a way that the answer provided directly responds to it.

    \item \textbf{Maintain Formality and Precision}: Use formal language and precise wording to ensure the question is professional and easy to understand.
\end{enumerate}

\noindent
Examples:

\bigskip
\noindent
\textbf{Example 1}

\noindent
\textbf{Answer:} \\
The PSLV-C56 mission is scheduled to be launched on Sunday, 30 July 2023 at 06:30 IST / 01:00 UTC. It will be launched from the Satish Dhawan Space Centre, Sriharikota, Andhra Pradesh, India.

\bigskip
\noindent
\textbf{Question:} \\
When is the scheduled launch date and time for the PSLV-C56 mission, and where will it be launched from?

\bigskip
\noindent
\textbf{Example 2}

\noindent
\textbf{Answer:} \\
The Great Wall of China, built over several dynasties, stretches approximately 13,170 miles. It was primarily constructed to protect Chinese states and empires from invasions and raids.

\bigskip
\noindent
\textbf{Question:} \\
What is the length of the Great Wall of China, and why was it primarily constructed?

\bigskip
\noindent
\textbf{Example 3}

\noindent
\textbf{Answer:} \\
Marie Curie was awarded the Nobel Prize in Physics in 1903 and the Nobel Prize in Chemistry in 1911 for her work on radioactivity and the discovery of radium and polonium.

\bigskip
\noindent
\textbf{Question:} \\
In which years did Marie Curie receive her Nobel Prizes, and for what contributions were they awarded?

\bigskip
\noindent
\textbf{Example 4}

\noindent
\textbf{Answer:} \\
The Amazon Rainforest, often referred to as the "lungs of the Earth," covers over 5.5 million square kilometers and spans across nine countries in South America.

\bigskip
\noindent
\textbf{Question:} \\
What is the total area of the Amazon Rainforest, and across how many countries does it span?

\bigskip
\noindent
Now, generate a question based on the given answer below:

\bigskip
\noindent
\textbf{Answer:} \\
\{answer\}

\bigskip
\noindent
\textbf{Question:}

\noindent
\rule{\linewidth}{0.3mm}

\noindent 
\rule{\linewidth}{0.3mm}
\begin{center}
\textbf{Enhanced Document Context for Cross-Encoder}
\end{center}

\noindent 
\rule{\linewidth}{0.3mm}
\textit{Question}: At about what age do adults normally begin to lose bone mass?

\textit{Enhanced Text}:  The actual answer to the following question is: Based on the given context, adults typically begin to lose bone mass around the age of 40. The key points are: - Bone mass reaches its peak during young adulthood, and then there is a slow but steady loss of bone beginning about age 40. - After about age 30, people can start to lose bone faster than their body makes it, which can weaken the bones and increase the risk of breakage. - The reduction of bone mass begins between ages 30 and 40, and continues to decline. So the summarized response is that adults normally begin to lose bone mass around the age of 40. 

The context (which refers to text that was used to answer this question) is: ['Age. There’s no way around it: loss of bone mass comes with age, laying the groundwork for low bone density and the potential of osteoporosis. We typically lose bone mass starting at age 40 and one in two women and one in four men over the age of 50 will fracture a bone at some point.', 'After about age 30, you can start to lose bone faster than your body makes it, which can weaken the bones and increase the risk of breakage. Some bone loss is natural as men and women age, but women are at higher risk of significant bone loss.', 'Bone mass reaches its peak during young adulthood. Then, after a period of stability, there is a slow but steady loss of bone beginning about age 40. In women, normal aging and menopause significantly increase susceptibility to osteoporosis.', 'In adults, this can take ten years. Until our mid-20s, bone density is still increasing. But at 35 bone loss begins as part of the natural ageing process. This becomes more rapid in post-menopausal women and can cause the bone-thinning condition osteoporosis.', 'The reduction of bone mass begins between ages 30 and 40, and continues to decline. Women lose about 8\% of skeletal mass every decade, while men lose about 3\%. Epiphyses, vertebrae, and the jaws lose more mass than other sites, resulting in fragile limbs, reduction in height, and loss of teeth.'].  

Answer Relevancy assesses how pertinent the actual answer is to the given context. It is measured between 0 and 1; where a lower score is given to answers that are incomplete or contain redundant information; and a higher score indicates better relevancy. For the given question, context and answer, the answer relevancy score is: 0.9531866263993314. 

Context Precision assesses how relevant is every context towards answering the question. Ideally all of the text in all of the contexts should be relevant to the question. It is measured betwen 0 and 1; where a lower score is given to lower precision contexts; and a higher score indicates more precision. For the given question and contexts, the context precision score is: 0.06666666666666667. 

Context recall measures the extent to which the context aligns with the ground truth. It is computed based on attributing text in the ground truth to the context, and is measured between 0 and 1; where a lower score is given to lower recall contexts; and a higher score indicates better performance. For the given ground truth and contexts, the context recall score is: 0.2727272727272727 

Faithfulness measures the factual consistency of the generated answer against the given context. It is considered faithful if all the claims that are made in the answer can be inferred from the given context. It is measured between 0 and 1; where a lower score is given to answers consisting of claims that are not in the context; and a higher score indicates that the answer is using information from the contexts. For the given question, context and answer, the faithfulness score is: 1.0.

\noindent
\rule{\linewidth}{0.3mm}

\noindent 
\rule{\linewidth}{0.3mm}
\begin{center}
\textbf{Synthetic Data Prompt Generation}
\end{center}

\noindent 
\rule{\linewidth}{0.3mm}

\noindent
\textbf{Prompt 1: Cloud Computation Sales and Marketing}

\bigskip
\noindent
\textbf{Task:}

\noindent
Generate a passage related to cloud computation sales and marketing and a corresponding question based on the passage.

\bigskip
\noindent
\textbf{Example:}

\noindent
\textbf{Passage:}

\noindent
Cloud computation has revolutionized sales and marketing by enabling businesses to analyze large datasets in real-time. This allows for more precise targeting of potential customers and more effective allocation of marketing resources. Companies can now leverage cloud-based tools to track consumer behavior, predict trends, and personalize marketing campaigns.

\bigskip
\noindent
\textbf{Question:}

\noindent
How has cloud computation changed the way businesses approach sales and marketing?

\bigskip
\noindent
\textbf{Prompt 2: Basketball}

\bigskip
\noindent
\textbf{Task:}

\noindent
Generate a passage related to basketball and a corresponding question based on the passage.

\bigskip
\noindent
\textbf{Example:}

\noindent
\textbf{Passage:}

\noindent
Basketball is a fast-paced sport that requires a combination of physical skill, strategic planning, and teamwork. Players must constantly communicate and adapt to the changing dynamics on the court. Successful teams often have a mix of strong defense, effective offense, and the ability to capitalize on opponents' mistakes.

\bigskip
\noindent
\textbf{Question:}

\noindent
What are the key components of a successful basketball team?

\bigskip
\noindent
\textbf{Prompt 3: Random Topics (Unrelated to Cloud Computation Sales and Marketing and Basketball)}

\bigskip
\noindent
\textbf{Task:}

\noindent
Generate a passage on a random topic unrelated to cloud computation sales and marketing or basketball and a corresponding question based on the passage.

\bigskip
\noindent
\textbf{Example:}

\noindent
\textbf{Passage:}

\noindent
The history of the automobile is marked by continuous innovation and technological advancements. From the invention of the internal combustion engine to the development of electric cars, the automotive industry has always been at the forefront of engineering and design. Modern cars are equipped with advanced safety features, autonomous driving capabilities, and environmentally friendly technologies.

\bigskip
\noindent
\textbf{Question:}

\noindent
How has the automotive industry evolved over the years?

\noindent 
\rule{\linewidth}{0.3mm}

\newpage
\begin{figure}[htbp]
    \centering
    \includegraphics[scale=0.8]{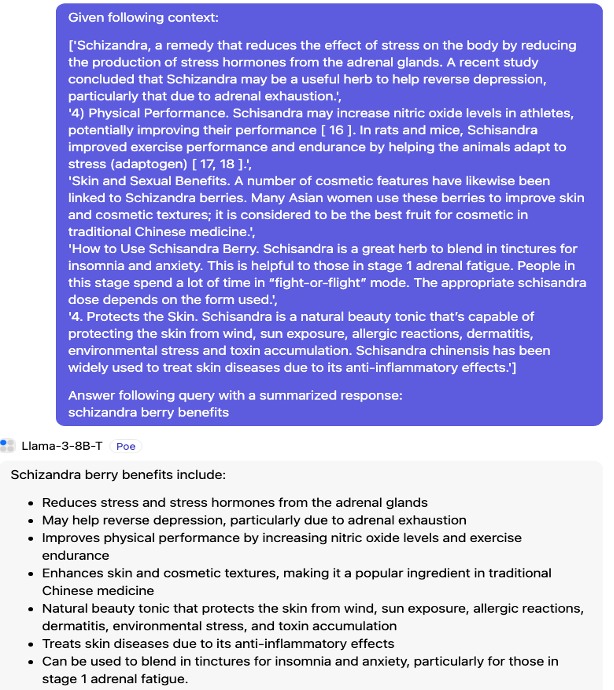}
    \label{fig:appredix2}
    \caption{Example of Prompt of RAG Summarization with Retrieved Chunks}
\end{figure}

\newpage

\subsection{Unbiased Estimator}
\label{sec:proof}
Let $f: X \xrightarrow{} Y$ be a black box function mapping from the input space $X$ (queries) to the output space $Y$ (LLM-based metrics space). For each query $x$, the output is a vector $y=(y_1,y_2,y_3,y_4)$ where $y_1$,$y_2$,$y_3$ and $y_4$ represent Retrieval Precision, Retrieval Recall, Faithfulness, and Answer Relevance, respectively.  

Assume a dataset $D = \{x_i\}_{i=1}^n$ where each $x_i$ is a query. Associated with each query is a vector of metrics $m_i= (m_{i1},m_{i2},m_{i3},m_{i4})$. For each metric $k$ (where $k = 1,2,3,4$ corresponding to the four metrics), the estimator $\hat{\theta}_k$ based on $n$ observations is given by the sample mean:
\[\hat{\theta}_k=\frac{1}{n}\sum_{i=1}^{n}m_{ik}\]
Based on the bootstrap procedure described in 3.2, for each bootstrap sample $D_b^*$  and for each metric $k$, compute the bootstrap replicate:
\[\hat{\theta}_k^*=\frac{1}{n}\sum_{i=1}^{n}m_{bik}^*\]
The bootstrap expectation for each metric $k$ over all bootstrap samples is:
\[E^* \bigr[ \hat{\theta}_k^* \bigr] =\frac{1}{B}\sum_{b=1}^{B}\hat{\theta}_k^*\]
We need to know that:
\[E^* \bigr[ \hat{\theta}_k^* \bigr] = \hat{\theta}_k \]

to prove the bootstrap statistics $\hat{\theta}_k^*$ are unbiased estimators of the empirical means $\hat{\theta}_k$ for each metric. 

\textbf{Proof:}
Given that each $\hat{\theta}_k^*$ is an average of $n$ independently and identically distributed (i.i.d.) bootstrap samples drawn with replacement from $m_k$, we apply the law of large numbers in the bootstrap world, stating: 
\[E^* \bigr[ \hat{\theta}_k^* \bigr] \approx \frac{1}{n}\sum_{i=1}^{n} E^* \bigr[ m_{ik}^*  \bigr] \]
Since the bootstrap samples are drawn from the empirical distribution of $m_k$, $E^* \bigr[ m_{ik}^* \bigr] = \hat{\theta}_k$. Therefore:
\[E^* \bigr[ \hat{\theta}_k^* \bigr] \approx \hat{\theta}_k \]

Thus, $\hat{\theta}_k^*$ is an unbiased estimator of $\hat{\theta}_k$ under the bootstrap distribution, assuming that the original sample is representative of the population and the metrics are i.i.d.

\end{document}